\documentclass[12pt,preprint]{aastex}

\begin{document}

\title{Planet formation around low mass stars: the moving snow line and super-Earths}
\shorttitle{THE MOVING SNOW LINE AND SUPER-EARTHS}

\author{Grant M. Kennedy}
\affil{Research School of Astronomy and Astrophysics, Mt Stromlo Observatory, Australian National
  University, Cotter Rd, Weston Ck, ACT 2611, Australia}
\affil{ANU Planetary Science Institute, Australian National University, Canberra, Australia}%
\email{grant@mso.anu.edu.au}

\author{Scott J. Kenyon}
\affil{Smithsonian Astrophysical Observatory, Cambridge, MA 02138, USA}
\email{kenyon@cfa.harvard.edu}

\author{Benjamin C. Bromley}
\affil{Department of Physics, University of Utah, 201 JFB, Salt Lake City, UT 84112}
\email{bromley@physics.utah.edu}

\begin{abstract}
  We develop a semi-analytic model for planet formation during the pre-main
  sequence contraction phase of a low mass star. During this evolution, the
  stellar magnetosphere maintains a fixed ratio between the inner disk radius
  and the stellar radius. As the star contracts at constant effective
  temperature, the `snow line', which separates regions of rocky planet
  formation from regions of icy planet formation, moves inward.  This process
  enables rapid formation of icy protoplanets that collide and merge into
  super-Earths before the star reaches the main sequence. The masses and orbits
  of these super-Earths are consistent with super-Earths detected in recent
  microlensing experiments.
\end{abstract}

\keywords{planetary systems: formation --- planetary systems: protoplanetary disks --- stars:
  evolution --- stars: formation}

\section{Introduction}

The recent discoveries of super-Earths around low mass stars challenge our
understanding of planet formation. From separate microlensing events,
\citet{2006Natur.439..437B} and \citet{2006ApJ...644L..37G} provide strong
evidence that planets with masses $M \sim$ 5--15\,$M_{\oplus}$ are common around
M dwarf stars\footnote{These microlensing observations directly yield the ratio
  $M_P/M_{\star}$, where $M_P$ is the mass of the planet and $M_{\star}$ is the
  mass of the star. Adopting the most likely parent star -- a K-dwarf or an
  M-dwarf -- yields the most likely $M_P$.}. With orbital semi-major axes $a
\sim 2.5$--3\,AU, these planets are probably ice giants roughly similar in
structure to Uranus and Neptune in the Solar System.

\citet{2006ApJ...644L..79B} proposes that these planets form in two stages.
After a disk instability produces a gas giant, photoevaporation of the gas 
giant atmosphere leaves behind an icy core with $M \sim$ 10--20 $M_{\oplus}$.  
This mechanism requires a massive disk to initiate the instability and a nearby
O-type star to photoevaporate the gas giant atmosphere.
\citeauthor{2006ApJ...644L..79B} notes that this process should yield (i)
super-Earths around M dwarfs formed in rich star clusters and (ii) gas giants
around M dwarfs formed in low mass stellar associations.

\citeauthor{2006Natur.439..437B} suggest that super-Earths favor coagulation
models, where collisions of 1--10 km objects eventually produce icy planets with
$M \sim$ 10 $M_{\oplus}$ at 1--10 AU. Although numerical calculations appear to
preclude gas giants at 1--10 AU around M dwarfs \citep{2004ApJ...612L..73L},
there has been no demonstration that coagulation produces icy planets on
reasonable timescales in a disk around an M dwarf.

Here, we develop a semi-analytic coagulation model, and show that contraction of
the central star along a pre-main sequence (PMS) Hayashi track sets the initial
conditions for planet formation around low mass stars.  Our results indicate
that icy protoplanets with $M \sim$ 0.1--1\,$M_{\oplus}$ form in $\sim$
0.1--1\,Myr at 1--4\,AU. Over 50--500\,Myr, collisions between protoplanets
produce super-Earths with masses similar to those detected in microlensing
surveys.

We start with the motivation for our study in \S \ref{sec:motivation}, discuss
the coagulation model of planet formation and the moving snow line in \S
\ref{sec:snowline}, develop the disk evolution model in \S \ref{sec:evolution},
and apply the model to super-Earth formation in \S \ref{sec:superEarths}. We end
with a brief summary in \S \ref{sec:summary}.

\section{Motivation: planet formation in the disk of a low-mass star}\label{sec:motivation}

To motivate our study, we contrast planet formation around low mass stars and
solar-type stars. For solar-type stars approaching the main sequence, the
luminosity is roughly constant on typical planet formation timescales of
10--100\,Myr. Thus, the conditions where planets form change little with time.
For stars with masses $\lesssim 0.5\,M_{\odot}$, however, the luminosity fades
by a factor of 10--100 on the Hayashi track. Because the inner disk radius is
`locked' at a fixed distance relative to the radius of the central star, the
inner disk contracts as the star contracts.  During this evolution, the `snow
line' -- the point that separates the inner region of rocky planet formation
from the outer region of icy planet formation -- also moves inward.

Coupled with the evolution of the inner disk, the moving snow line produces a
dramatic variation in the surface density at fixed distances from the central
star. This behavior enables the rapid formation of icy protoplanets.  As the low
mass star approaches the main sequence, these protoplanets collide and merge
into super-Earths with properties similar to those detected in recent 
microlensing experiments.

\section{Coagulation and the Moving Snow Line}\label{sec:snowline}

In coagulation models, planets grow from repeated collisions and mergers of
small objects in a circumstellar disk \citep{1969QB981.S26......}. When 1--10~km
`planetesimals' form and start to grow
\citep{1980Icar...44..172W,2005A&A...434..971D}, dynamical friction damps the
orbital eccentricities of the largest objects.  Damping yields large
gravitational cross-sections and leads to `runaway growth,' where the largest
objects grow fastest and run away from more slowly growing smaller objects
\citep{1989Icar...77..330W,1996Icar..123..180K}. Throughout the runaway, the
largest protoplanets stir up the leftover planetesimals. Eventually, the
leftovers have orbital velocity dispersions comparable to the escape velocities
of the largest protoplanets.  Because gravitational cross-sections fall as
velocity dispersions rise, runaway growth ends. The ensemble of planetesimals
and protoplanets then enters `oligarchic' growth, where the largest objects --
oligarchs -- accrete at rates roughly independent of their size
\citep{1998Icar..131..171K}.

During oligarchic growth, protoplanets become isolated from their surroundings.
If an oligarch accretes all of the mass in an annulus with width $2 B R_H$,
where $R_H = a (M / 3 M_{\star})^{1/3}$ is the Hill radius, its isolation mass
is
\begin{equation}\label{eq:miso}
  M_{iso} \approx 4 \pi a B R_H \sigma 
  \propto (B \sigma)^{3/2} a^3 M_{\star}^{-1/2} ~,
\end{equation}
where $a$ is the orbital semi-major axis and $\sigma$ is the mass surface
density of solid material in the disk
\citep[e.g.][]{1993ARA&A..31..129L,2000Icar..143...15K}.  If $\rho$ is the mass
density of a solid object the timescale to reach isolation is
\citep{2004ApJ...614..497G}
\begin{equation}\label{eq:tiso}
  t_{iso} \propto \rho^{1/2} a^{3/2} \sigma^{-1/2} ~.
\end{equation}
In the Solar System, oligarchic growth has two regimes. For $a \lesssim 3$\,AU,
planetesimals are rocky because volatile materials remain in the gas. In the
`Minimum Mass Solar Nebula' \citep[MMSN,
][]{1977Ap&SS..51..153W,1981PThPS..70...35H} with $B~=~2.5-5$, $\rho \sim
3$\,g\,cm$^{-3}$, and $\sigma \sim 8$\,g\,cm$^{-2}$ at 1\,AU, $M_{iso}
\sim0.05$--0.1\,$M_{\oplus}$.  Once oligarchs contain $\sim$\,50\% of the total
mass in solids, their mutual dynamical interactions lead to `chaotic' growth
\citep{2004ApJ...614..497G,2006AJ....131.1837K}, where collisions between
oligarchs eventually produce Earth-mass planets. Numerical simulations suggest
that $\sim$10--20 oligarchs collide to form a typical Earth-mass planet in
$\sim$10--100 Myr at 1\,AU around a solar-type star
\citep{2001Icar..152..205C,2004Icar..168....1R,2006AJ....131.1837K}.

Outside the `snow line', ice condensation enhances $\sigma$ and promotes the
formation of larger oligarchs.  For $\rho \sim 1.5$\,g\,cm$^{-3}$ and $\sigma
\sim 3$--6\,g\,cm$^{-2}$ at 5\,AU, isolated oligarchs with $M_{iso} \sim
5\,M_{\oplus}$ form on timescales $t_{iso} \sim 1$--3\,Myr.  These icy oligarchs
accrete gas directly from the nebula and grow into gas giant planets in several
Myr \citep{1996Icar..124...62P}, comparable to the lifetime of the gaseous disk
\citep[e.g.][]{2000prpl.conf..401H,2001ApJ...553L.153H,2004ApJS..154..428Y,2005AJ....129..935C}.

For solar-type stars, planet formation is fairly independent of stellar
evolution. Throughout most of the PMS phase, the solar luminosity is roughly
constant. Thus, the position of the snow line -- $a_{snow} \sim
(L_{\star}/T_{snow}^4)^{1/2}$, where $L_{\star}$ is the stellar luminosity and
$T_{snow}$ is the temperature where water and other volatile materials condense
out of the gas -- is roughly stationary in time. Because the $\sim$0.1--1\,Myr
formation time for planetesimals and oligarchs is short compared to the
$\sim$10\,Myr PMS lifetime, the separation between icy and rocky (proto)planets
remains fairly distinct, evident in the composition of different populations in
the asteroid belt \citep{2000orem.book..413A,2002aste.conf..235R}. Although
there is some mixing between water-rich and water-poor regions
\citep{2004Icar..168....1R}, most of chaotic growth occurs when the Sun lies
close to the main sequence at nearly constant $L_{\star}$.

In contrast with solar-type stars, stellar evolution is a crucial feature that
defines the nature of newly-formed planets around low mass stars. Because the
timescale for planetesimal and oligarch formation is short compared to the
0.1--1\,Gyr PMS contraction time
\citep{1994ApJS...90..467D,1998A&A...337..403B,2000A&A...358..593S}, the timing
of planetesimal formation sets the nature of icy/rocky planets with distance
from a low mass star.  On its Hayashi track, the luminosity of a
0.25\,$M_{\odot}$ star fades by a factor of several hundred at roughly constant
effective temperature.  During this period, $a_{snow}$ moves inward by a factor
of $\sim$15--20.  Just outside the moving snow line, ice condensation increases
$\sigma$ ($M_{iso}$) by a factor of $\sim$ 4 (8)
\citep{1981PThPS..70...35H}; $t_{iso}$ decreases by a factor of 3.
This moving snow line enables rapid formation of icy oligarchs that can
collide and merge into super-Earths.

\section{Evolution of a disk around a contracting star}\label{sec:evolution}

Disk evolution is also an important feature of planet formation around low mass
stars. In the standard MMSN model, $\sigma$ is fixed in time and scales with the
stellar radius on the main sequence \citep[e.g.][]{1981PThPS..70...35H}.
However, when PMS stars actively accrete from a circumstellar disk, magnetic
interactions between the star and the disk appear to `lock' the inner disk
radius $R_{in}$ at a fixed distance relative to the stellar radius, $\xi \equiv
R_{in}/R_\star \sim 3$, at several Myr \citep[e.g.][]{2005ApJ...623..952E}.
Although the duration of this phase is not well-constrained, the observed change
in $\xi$ for disks around solar-type stars is a factor of $\sim 2$--3
\citep{2005ApJ...623..952E}.  If disks around low mass stars remain locked for
the entire PMS phase, the maximum decrease in the inner disk radius is a factor
of $\sim$15--20. This change is much larger than the observed variation of
$\xi$; thus we assume $\xi$ = constant.  To conserve mass and angular momentum,
$\sigma$ and the outer disk radius must evolve, which impacts $M_{iso}$ and the
formation timescales for oligarchs and planets.

To construct a model for disk evolution, we adopt
\begin{equation}\label{eq:sigma}
  \sigma(t) = \sigma_0 \frac{ M_{\star} }{ M_\odot } f_{ice}
  \left( \frac{ R_\star(t) }{\beta \, a_{\textrm{{\tiny AU}}}} \right)^{3/2}
\end{equation}
where $R_\star$ is in units of solar radii, $\sigma_0 = 8$\,g\,cm$^{-2}$, and
$a_{{\textrm{{\tiny AU}}}}$ is the radial distance from the star in AU. Setting
the scale factor $\beta \sim 3$ yields the usual $\sigma$(MMSN) for a
1\,$M_{\odot}$ star at 1\,Myr, when a large fraction of the solid mass in the
terrestrial zone of the Solar System is in large bodies. Consistent with
observations \citep{2000prpl.conf..559N,2006ApJ...645.1498S}, we scale $\sigma$
and the disk mass linearly with the stellar mass.  For a 1\,Myr old
$0.25\,M_\odot$ star, this scaled MMSN has $\beta = 2$ and $M_{disk} = 0.026
M_\star$ integrated from $3 R_\star$ to 50\,AU for a gas/solids ratio of 100.
To provide a smooth transition from $f_{ice} = 1$ for $a \lesssim a_{snow}$ to
$f_{ice} = 4$ for $a \gtrsim a_{snow}$ \citep{1981PThPS..70...35H}, we include a
parameter $f_{ice}=1+(\Delta_{ice}-1)/(1+e^x)$ where $\Delta_{ice}=4$,
$x=(a_{snow}-a)/\Delta T_{snow}(a)$ and $\Delta T_{snow}(a)$ is the radial
distance equivalent to a 5\,K temperature change.

To derive $a_{snow}$, we adopt the temperature profile of a flat circumstellar
disk, $T \propto T_\star \left( R_\star/a \right)^{3/4}$
\citep{1987ApJ...323..714K}.  We scale this relation to place the snow line at
2.7\,AU at 1\,Myr for a $1\,M_{\odot}$ mass star, as inferred from analyses of
water-rich objects in the outer asteroid belt
\citep{2000orem.book..413A,2002aste.conf..235R}. To evaluate
$L_{\star}(t),R_\star(t),$ and $T_\star(t)$, we use PMS evolutionary tracks from
\citet{2000A&A...358..593S}; other tracks
\citep{1994ApJS...90..467D,1998A&A...337..403B} yield similar results.

With these ingredients, we derive the evolution of $\sigma$, $M_{iso}$, and
$t_{iso}$ as the star contracts to the main sequence.  This evolution has two
main features. Initially, the snow line is at a large distance, $a_{snow} \sim
5$\,AU, from the luminous PMS star. Inside 1--2\,AU, rocky oligarchs form and
reach $M_{iso}$ before the star contracts significantly.  Outside 1--2\,AU,
$t_{iso}$ is long (eq.\,\ref{eq:tiso}) compared to the initial contraction time.
As the star contracts, ices condense out of the nebula and the snow line moves
inward.  For $a \lesssim 1$--2\,AU, this material coats the growing oligarchs,
leftover planetesimals, and the surrounding debris with an icy veneer that may
extend the oligarchic growth phase and produce more massive oligarchs. For $a
\gtrsim 1$--2\,AU, ice condensation reduces $t_{iso}$ by a factor $\sim$ 3
(eq.\,\ref{eq:tiso}), which enables the rapid formation of icy oligarchs well
before the central star reaches the main sequence.

\section{Super-Earth Formation}\label{sec:superEarths}

\begin{figure}
  \plotone{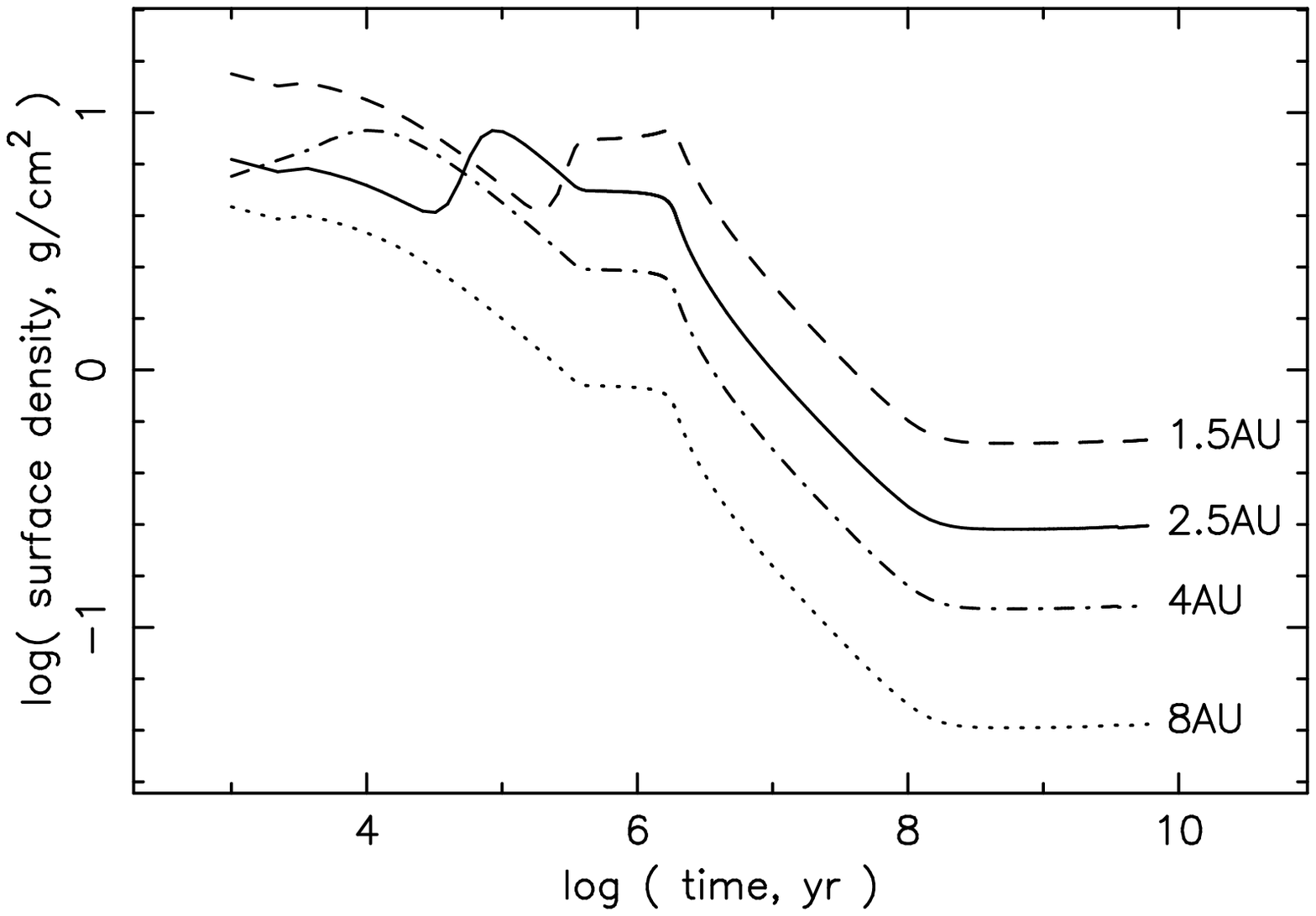}
  \caption{Surface density evolution at fixed radii around a $0.25\,M_\odot$
    star with $M_{disk}/M_\star = 0.065$. As the snow line moves inwards, ice
    condensation increases $\sigma$, which leads to faster formation of more
    massive oligarchs.}\label{fig:sigma_a_paper.ps}
\end{figure}

To explore the consequences of this picture, we consider a disk with $\beta \sim
1$ ($M_{disk}/M_\star = 0.065$), which lies at the upper end of the range
inferred from observations\footnote{In their coagulation model for Neptune,
  \citet{2004ApJ...614..497G} also consider a disk with $M_{disk} \sim
  3$--6\,$M_{MMSN}$.}
\citep{1995ApJ...439..288O,1997A&A...324.1036N,1998A&A...330..549N,2000prpl.conf..559N,2006ApJ...645.1498S}.
Figure \ref{fig:sigma_a_paper.ps} shows the $\sigma$ evolution for this system
at several distances from a $0.25\,M_\odot$ star. For disks with other masses,
$\sigma \propto M_{disk}$, $M_{iso} \propto M_{disk}^{3/2}$, and $t_{iso}
\propto M_{disk}^{-1/2}$.  Aside from the long-term decline in $\sigma(a)$ from
PMS evolution, the $\sigma$ evolution shows clear increases when the snow line
crosses specific points in space and ices condense out of the gas. At these
times, $\sigma$ remains at a relatively constant plateau value for
$\sim$1--3\,Myr before declining monotonically as the central star approaches
the main sequence.

During the plateau phases, the timescale for oligarchs to reach $M_{iso}$
($\lesssim 1$\,Myr; eq.\,\ref{eq:tiso}) is shorter than the stellar contraction
time. Because the isolation masses at fixed distances decrease as $\sigma$
decreases (eq.\,\ref{eq:miso}), these early times provide the best opportunity
for coagulation to form large protoplanets.

\begin{figure}
  \plotone{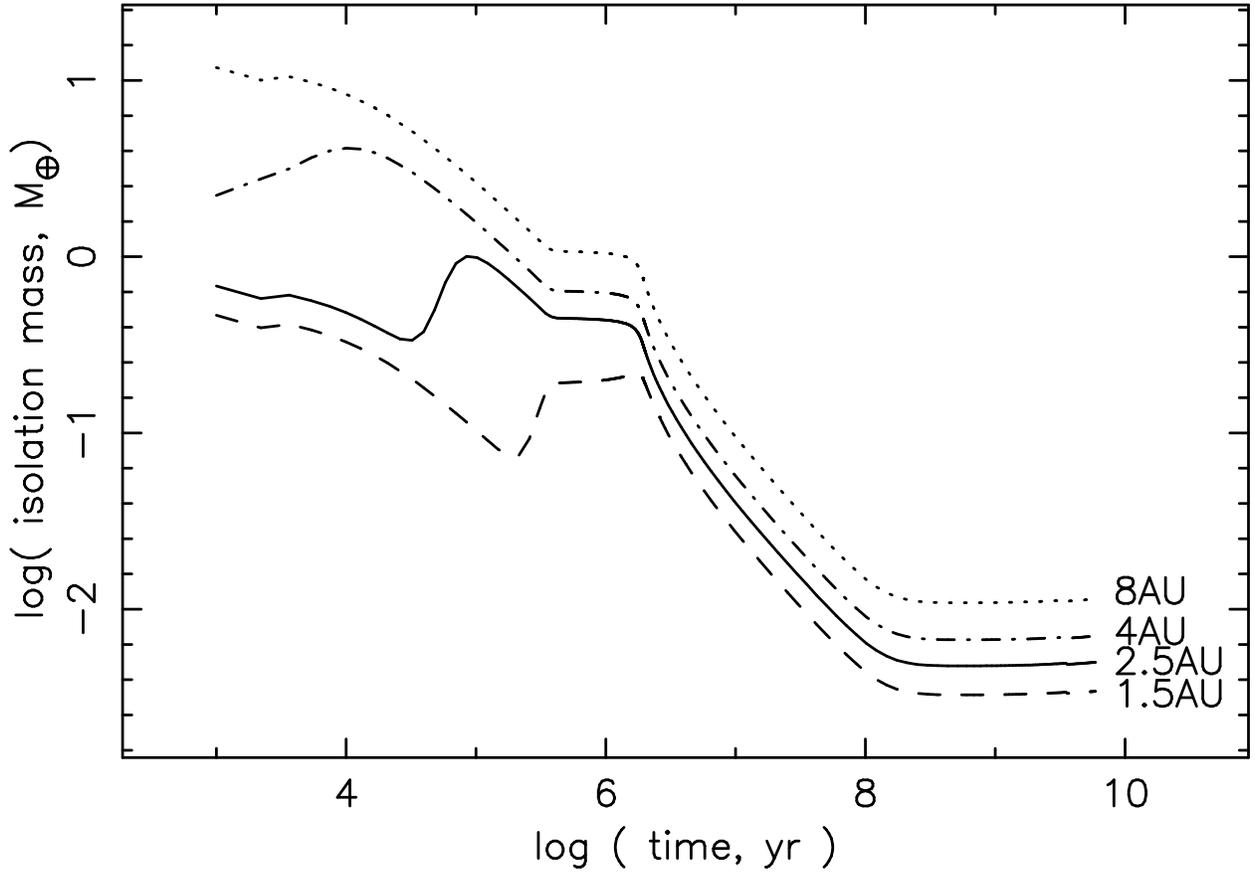}
  \caption{Evolution of $M_{iso}$ at fixed radii using the $\sigma$ evolution
    of Figure\,\ref{fig:sigma_a_paper.ps}. Ice condensation leads to more
    massive oligarchs at 2--8 AU in 1 Myr.}\label{fig:miso_a_paper.ps}
\end{figure}

Figure \ref{fig:miso_a_paper.ps} shows the time evolution of $M_{iso}$.
Interior to $a_{snow}$ ($a \lesssim 1$--2\,AU), rocky oligarchs with $M_{iso}
\sim 0.1\,M_\oplus$ form in $\sim$10$^5$\,yr.  As the star contracts, ice
condensation enables the formation of larger oligarchs with $M_{iso} \sim
0.2\,M_\oplus$ in $\sim$1\,Myr.  At $a \sim 2$--3\,AU, ice condensation during
runaway growth promotes the formation of oligarchs with $M_{iso} \sim
0.5\,M_\oplus$ in $\sim 10^5$\,yr.

This analytic prescription for protoplanet growth suggests that oligarchs with
$M_{iso} \sim 0.1$--0.5\,$M_{\oplus}$ can form at $\sim$1--3\,AU in
$\lesssim$\,1\,Myr. If oligarchs contain roughly half the mass in solid material
at the onset of chaotic growth, our model disk with $\sim5$--10\,$M_{\oplus}$ at
1--4 AU will have $\sim$10--100 oligarchs.  The model predicts $\sim$10
oligarchs at 2--4\,AU. Thus, the building blocks for observable super-Earths can
form on timescales much shorter than disk lifetimes derived from measurements of
dust emission from low mass PMS stars
\citep{2002AJ....124..514S,2004AJ....127.2246W,2004ApJ...608..526L,2005ApJ...631.1161P}.

To consider whether oligarchs can merge into super-Earths on reasonable
timescales, we follow \citet{2004ApJ...614..497G} and introduce a parameter
$\mathcal{R} = v_{esc}/\Omega a$, where $v_{esc}$ is the escape velocity and
$\Omega a$ is the orbital velocity of a growing planet. When $\mathcal{R} \ll$
1, colliding oligarchs merge; when $\mathcal{R} \gg$ 1, collisions often eject
an oligarch from the planetary system. At 1--5\,AU around a $0.25\,M_{\odot}$
star, this merger condition ($\mathcal{R} \lesssim$ 1) allows the formation of
$\sim 5\,M_\oplus$ planets at 1--2\,AU, $\sim 3\,M_\oplus$ planets at 2--3\,AU,
and 1--2\,$M_\oplus$ planets at 3--4\,AU. In this analytic model, the timescale
to produce planets is $\sim 200$\,Myr (1\,Gyr) at 1\,AU (5\,AU).

To derive another estimate for the masses and formation timescales, we consider
the results of complete numerical simulations of planet formation from an
initial ensemble of oligarchs \citep{2001Icar..152..205C,2004Icar..168....1R} or
planetesimals \citep{2006AJ....131.1837K}. In the solar terrestrial zone,
collisions and mergers of 10--20 oligarchs with masses $\sim M_{iso}$ yield 2--5
planets with masses comparable to the mass of the Earth on timescales of
10--100\,Myr. Although the final orbital parameters depend on the late-time
evolution of the planetesimals and the gaseous disk, the typical masses and
collision histories of Earth-mass planets are similar in all calculations and
agree fairly well with analytic estimates.  Adapting this collisional history to
a planetesimal disk around a 0.25 $M_{\odot}$ star, mergers of $\sim$ 10
oligarchs should yield planets with masses $\sim$1--2\,$M_{\oplus}$ at 1\,AU and
$\sim 3$--5\,$M_{\oplus}$ at 2.5\,AU.

Combining the analytic and scaled numerical results, the timescale for oligarchs
to merge into planets is roughly
\begin{equation}\label{eq:tcoag}
  t_{merge} \sim 10-100 \left( \frac{\rm 8~g~cm^{-2}}{\sigma} \right )
  \left ( \frac{P}{\rm 1~yr} \right ) \, {\rm Myr} \,\,
\end{equation}
where $P$ is the orbital period. Thus, the expected merger timescale for
oligarchs at 1--3\,AU around a $0.25\,M_\odot$ star is $\sim$2--5 times longer
than for the terrestrial zone around a solar-type star. This timescale is
comparable to the duration of the PMS contraction phase and is much shorter than
the expected stellar lifetime.  Thus, coagulation can produce super-Earths
around low mass stars on timescales of $\sim 50$--500\,Myr.

In constructing our model, we adopted a standard surface density law, $\sigma
\propto a^{-3/2}$, and ignored details of the disk structure
\citep[e.g.][]{2005ApJ...620..994D,2006ApJ...640.1115L} and physical mechanisms
for ice condensation \citep[e.g.][]{2004M&PS...39.1859P}.  Although details of
the disk structure and ice condensation mechanisms can affect the position of
the snow line, our main conclusions that (i) $a_{snow}$ moves considerably
during the PMS contraction of a low mass star, and (ii) ice condensation during
the PMS contraction phase produces massive oligarchs in $\sim 0.1$--1\,Myr and
super-Earths in $\sim$ 100 Myr, are generally independent of these details. The
main uncertainties in our picture are the probability of the large initial disk
mass and the details of the final accretion stage when 1--$2\,M_{\oplus}$
planets evolve into 3--$5\,M_{\oplus}$ planets. Observations of larger samples
can yield better estimates for the range of initial disk masses for low mass
stars and for the $M_{disk}$--$M_{\star}$ relation.  Detailed numerical
simulations can provide better estimates of the masses and formation timescales
for super-Earths.

\section{Model summary and predictions}\label{sec:summary}

We have developed an analytic prescription for planet formation by coagulation
around low mass stars. The model has two distinctive features that enable
formation of super-Earths during the PMS contraction phase.

\begin{itemize}

\item We set the inner disk radius at a fixed distance relative to the radius of
  the central star, $\xi \equiv R_{in}/R_{\star}$. Thus, $R_{in}$ changes as the
  star contracts to the main-sequence, leading to significant evolution in
  $\sigma(a)$.

\item During PMS contraction, $a_{snow}$ moves inwards by a factor of
  $\sim$15--20, producing large enhancements in $\sigma(a)$ as ices condense out
  of the nebula. Ice condensation is the key mechanism that allows coagulation
  to produce super-Earths around low mass stars. This process results in new
  outcomes for planet formation, including planets with dense, rocky cores and
  thick, icy surfaces.

\end{itemize}

We applied this model to super-Earth formation around a $0.25\,M_{\odot}$ star.
At 1--5\,AU, isolated oligarchs can grow to masses $\sim 0.1$--1\,$M_{\oplus}$ in
$\sim0.1$--1\,Myr, short compared to the $\sim 100$\,Myr PMS contraction time.
These oligarchs merge into super-Earths with masses $\sim 2$--5\,$M_{\oplus}$ as
the star contracts to the main sequence. Thus, coagulation can produce planetary
systems similar to those detected in recent microlensing events.

Aside from our success in producing icy super-Earths at 1--3\,AU around low mass
stars, the model makes clear predictions for rocky planet formation close to a
low mass star.  At $a \lesssim 1$\,AU, the isolation masses estimated for rocky
planets are $M_{iso} \lesssim 0.01\,M_{\oplus}$. If $\sim$10 isolated objects
merge into a rocky planet, we predict many low mass planets with masses
$\sim$0.1\,$M_{\oplus}$ at distances of 0.05--0.5\,AU around low mass M dwarfs.
We plan to explore the consequences of our model in future papers.

\acknowledgements We acknowledge support from the {\it NASA Astrophysics Theory
  Program} through grant NAG5-13278 (SK, BB) and an Australian Postgraduate
Award (GK).  We thank T. Currie, M. Geller, the ANU Planetary Science Institute
planet group, and an anonymous referee for helpful comments on the project and
the manuscript.

\clearpage

\end{document}